\documentstyle[aps,manuscript]{revtex}
\begin{document}
\draft
\title{Complex space-time and the classification of elementary particles}
\author{Lu Lin}
\address{Department of Electrophysics, National Chiao Tung\\
University, 1001 Ta Hsueh Road, Hsinchu, 30050 Taiwan}
\maketitle

\begin{abstract}
It is shown that the Dirac theory implies complex space-time and complex
space-time can lead to the Dirac equation. \ It is suggested that fermions
are grouped into doublets, those doublets are then divided into color
singlets (leptons) and color triplets (quarks), then they are further
divided into generation singlets and generation triplets. \ If the exclusion
principle for fermions works in the internal space, then the non-observation
of free quarks can be explained. \ However, it is suggested that a possible
new free quark may exist with a color triplet and generation singlet state
which is antisymmetric in the internal space.

\noindent 
\end{abstract}


%\begin{center}

%\end{center}

\preprint{Preprint submitted to Physical Review B } \vspace{3mm} 
\narrowtext

\section{Introduction}

In his great discovery of the Dirac equation [1,2], Dirac introduced $\gamma
_{\mu }$ into the ory and describes naturally internal spin and
antiparticles. However, there are many more internal variables such as
isospin, hypercharge, color, ... etc. To have a theory that contains all of
these variables, new degrees of freedom for the internal motion may be
needed. The purpose of this article is to show that Dirac theory implies the
existence of a complex space-time, that complex space-time can account for
the fact that a free particle is a free wave as well, and also can reproduce
the Dirac equation. It is suggested that all fermions can be classified into
fermion doublets, color singlets and triplets, and generation singlets and
triplets. Finally, if the principle of anti-symmetrization for fermion
state, can be generalized to be valid in the internal space, then the
non-observation of a free quark is a result of the present treatment.
However, it is also suggested that there is a possibility that a new free
quark may exist with an internally anti-symmetric state function.

\section{Dirac theory and complex space-time}

Consider the Dirac Hamiltonian of a free election

\begin{equation}
H=\overrightarrow{\alpha }\cdot \overrightarrow{P}+\beta m,\text{ \ }%
(c=\hbar =1).  \eqnum{1}
\end{equation}
The $x_{1}$-component of the velocity and the $x_{1}$-component of
coordinate are given as(see Chapter XI of ref.[2])

\begin{equation}
\dot{x}_{1}=\alpha _{1}=P_{1}H^{-1}+\frac{1}{2}i\alpha _{1}^{0}\exp
(-2iHt)H^{-1}\equiv \dot{x}_{1}(classical)+\dot{\eta}_{1},  \eqnum{2}
\end{equation}

\begin{equation}
x_{1}=p_{1}H^{-1}t+a_{1}-\frac{1}{4}\dot{\alpha}_{1}^{0}\exp (-2iHt)H\
\equiv x_{1}(classical)+\eta _{1},  \eqnum{3}
\end{equation}
with $x_{1\text{ }}$(classical) $=a_{1}+p_{1}H^{-1}t,$ $\dot{x}_{1}$%
(classical) $=p_{1}H^{-1}$. We can consider the system of a particle with $%
(x_{\mu },p_{\mu };\mu =1,\cdots 4)$ in the physical space coupled to its
internal variables $(\alpha _{i},\beta ;i=1,2,3)$ in the internal space. Due
to the coupling, the generalized normal coordinates become mixtures of both
parts (the physical part and the internal part), and similarly for the
velocity components. The motion of the $\eta _{i}$ is known as
Zitterbewegung which has never been observed experimentally, so it cannot
exist in our physical space, and can only stay in the internal space. In
fact, $\eta _{i}$ is related to the $\alpha _{i}$ and $\beta $. Since the
total angular momentum is conserved

\begin{equation}
\frac{d}{dt}(\overrightarrow{r}(classical)\times \overrightarrow{P}+%
\overrightarrow{\eta }\times \overrightarrow{P}+\frac{1}{2}\overrightarrow{%
\sigma })=\frac{d}{dt}(\overrightarrow{\eta }\times \overrightarrow{P}+\frac{%
1}{2}\overrightarrow{\sigma })=0.  \eqnum{4}
\end{equation}
$\overrightarrow{{\bf \eta }}${\bf \ }is responsible for the conservation of
total angular momentum (orbital plus spin). From equation (2), $\dot{\eta}%
_{1}$ is a part of $\alpha _{1}(\dot{\eta}_{i}=\alpha _{i}$ when $%
\overrightarrow{P}$ is zero). This is because $\overrightarrow{\alpha } $ is
responsible for the Lorentz covariant 4-currents which are physically very
important in describing interactions between quantum systems. Therefore $%
\overrightarrow{{\bf \eta }}$ is physically important. A good physical
quantity must be Lorentz covariant, so there must exist an $\eta _{4\text{ }%
} $so as to make $\eta _{\mu }$ Lorentz covariant. Consider the case when $%
P_{i}=0$, and $\eta _{i}=(-i/2m)\beta \alpha _{i}$, \ so $\eta _{\mu
}=(-i/2m)\gamma ^{\mu }$. \ $\eta _{\mu }$ is a component of a dynamical
quantity in the internal space implying that there are 4 axes, $\ y_{\mu }$
in that space. The $\eta _{\mu }$ is complex. In our physical space-time, $%
x_{i}$ is real and $x_{4\text{ }}$is imaginary. There is no a priori reason
why the $y_{i\text{ }}$or $y_{4\text{ }}$should be real or imaginary when
the universe was created. However, nature has preference for simplicity and
unification. Therefore, the physical space-time and the internal space
should be created together as a whole by taking all possible combinations of 
$x_{i}$ and $x_{4\text{ }}$with equal probability to make a greater universe
as

\begin{equation}
GU=P_{1}(x_{i}^{R},x_{4}^{R})+P_{2}(x_{i}^{R},x_{4}^{I})+P_{3}(x_{i}^{I},x_{4}^{R})+P_{4}(x_{i}^{I},x_{4}^{I}),
\eqnum{5}
\end{equation}
where R=real, I= imaginary. The $P_{2}$ part is the physical space-time and
the other three parts are the internal space. A 4-vector in this GU is in
general complex and we conclude that the Dirac theory implies a complex
space-time. (In 1998, the author [3] suggested a complex space-time,
however, the derivation was not so satisfactory).

To describe a free particle with complex space-time, there must exist a
well-behaved physical function $\Phi $. Suppose we try a harmonic function
that has continuous partial derivatives of the second order and that
satisfies the Laplace equation. For a stationary state, the spatial
distribution must be independent of time, so $\Phi $ can be factorized and
we have

\begin{equation}
\Phi (X,T)=\Phi (X)\Phi (T),\text{ \ \ }X=x+i\overline{x},\text{ \ \ }T=T+i%
\overline{t},  \eqnum{6}
\end{equation}

\begin{equation}
\frac{\partial ^{2}\Phi (X)}{\partial x^{2}}+\frac{\partial ^{2}(X)}{%
\partial \overline{x}^{2}}=0,\text{ \ \ }\frac{\partial ^{2}\Phi (T)}{%
\partial t^{2}}+\frac{\partial ^{2}\Phi (T)}{\partial \overline{t}^{2}}=0. 
\eqnum{7}
\end{equation}
Since ${\bf x}$ and $\overline{x}$ are basically independent, we can write $%
\Phi (X)=\Phi (x)\Phi (\overline{x})$, and similarly for $\Phi (T)$. The
solution can be written as

\begin{equation}
\Phi =A\exp (\pm i\omega t\pm ik_{x})\cdot \exp (\pm \omega \overline{t}\pm k%
\overline{x}).  \eqnum{8}
\end{equation}
We can choose the direction of $k$ to be in the $x\pm i\overline{x}$ axes.
Note that there is a non-physical solution where $\omega $ and $k$ are
changed into $i\omega $ and $ik$ in equation (8). Setting $\overline{x}=%
\overline{t}=0$ , we get a projected solution with only real space-time
variables. The result is a free wave with $\omega $ and $k$ as the frequency
and wave number. That is, a free particle is a free wave as well.

Consider $\Phi (T)$ of equation (8) in the rest frame with the wave number $%
k=0$. The magnitude of $\omega $ is uniquely determined by the rest mass of
the particle and its value can assume to be $\pm \omega $ and $\pm i\omega $%
. If we require $\Phi $ to remain finite for all values of $t$ and $%
\overline{t}$, there are four independent wave functions in the complex
T-plane given as

\begin{equation}
\phi _{1}=e^{-i\omega t-\omega \overline{t}},\text{ \ \ }\phi
_{2}=e^{-\omega t-t\omega \overline{t}},\text{ \ \ }\phi _{3}=e^{i\omega
t-\omega \overline{t}},\text{ \ \ }\phi _{4}=e^{-\omega t+i\omega \overline{t%
}},  \eqnum{9}
\end{equation}
(set $t=\overline{t}=0$ at the moment of the creation of the universe). If $%
t $ and $\overline{t}$ are exchanged, $\phi _{1}$ and $\phi _{2}$ would be
exchanged and also for $\phi _{3}$ and $\phi _{4}$. Therefor, there is a
real-imaginary symmetry in the system. Define the generalized energy
operator as

\begin{equation}
\widehat{E}=\widehat{E}_{R}+\widehat{E}_{1}=i\left( \frac{\partial }{%
\partial t}+\frac{\partial }{\partial t}\right) .  \eqnum{10}
\end{equation}
It is easy to see that $\phi _{1}$ and $\phi _{2}$ have the same generalized
energy. This can be viewed as the system having an internal degree of
freedom with two allowed states with the same generalized energy in the
complex T-plane. We therefore have an SU(2) internal symmetry. Consider an
SU(2) transformation U

\begin{equation}
\left( 
\begin{array}{l}
\varphi _{1} \\ 
\varphi _{2}
\end{array}
\right) =U\left( 
\begin{array}{l}
\phi _{1} \\ 
\phi _{2}
\end{array}
\right) =\left( 
\begin{array}{l}
a_{1}\phi _{1}+a_{2}\phi _{2} \\ 
b_{1}\phi _{1}+b_{2}\phi _{2}
\end{array}
\right) .  \eqnum{11}
\end{equation}
The generalized energy of $(\varphi _{1},\varphi _{2})$ does not change.
That is, the generalized energy of $(\varphi _{1},\varphi _{2})$ is
invariant under an SU(2) transformation, and $(\varphi _{1},\varphi _{2})$
forms a basis represention of an SU(2) group which has the three $\sigma
_{i} $ as generators. Similarly, it is also true for the negative frequency
solutions $(\phi _{3},\phi _{4})$. By combining the four functions together
we can write

\begin{equation}
\psi _{1}=\varphi _{1}\left( 
\begin{array}{l}
1 \\ 
0 \\ 
0 \\ 
0
\end{array}
\right) ,\psi _{2}=\varphi _{2}\left( 
\begin{array}{l}
0 \\ 
1 \\ 
0 \\ 
0
\end{array}
\right) ,\psi _{3}=\varphi _{3}\left( 
\begin{array}{l}
0 \\ 
0 \\ 
1 \\ 
0
\end{array}
\right) ,\psi _{4}=\varphi _{4}\left( 
\begin{array}{l}
0 \\ 
0 \\ 
0 \\ 
1
\end{array}
\right) ,  \eqnum{12}
\end{equation}
with $\varphi _{1}=a_{1}\phi _{1}+a_{2}\phi _{2},$ $\varphi _{2}=b_{1}\phi
_{1}+b_{2}\phi _{2},$ $\varphi _{3}=a_{1}^{\prime }\phi _{3}+a_{2}^{\prime
}\phi _{4},$ $\varphi _{4}=b_{1}^{\prime }\phi _{3}+b_{2}^{\prime }\phi _{4}$%
. To get physical solutions in the rest frame with only real time $t$, we
project out the imaginery time by putting $\overline{t}=0$ in equations (9)
and (12). For a stationary state, we must also project out the unphysical
part by setting $a_{2}=b_{2}=a_{2}^{\prime }=b_{2}^{\prime }=0$, otherwist
the state is a mixture of $e^{-i\omega t}$and $e^{-\omega t}$. The four
resulting functions form a complete set of basis functions of the time $t$
with internal symmetry. Any stationary state function of the particle can be
written as

\begin{equation}
\psi (\overrightarrow{r},t)=\psi (\overrightarrow{r})\left[ \sum A_{j}\psi
_{j}\text{(projected)}\right] .  \eqnum{13}
\end{equation}
If $\ \psi (\overrightarrow{r},t)$ of equation (13) is a proper state
function of the system, there must exist a conserved current which is
Lorentz covariant and satisfies the continuity condition. Noticing that each 
$\psi _{i}$ is a spinor function, it is natural to construct a spinor
current $j^{\mu }$ as

\begin{equation}
j^{\mu }=\overline{\psi }\gamma ^{\mu }\psi ,\text{ \ }\overline{\psi }=\psi
^{\dagger }\gamma ^{0},  \eqnum{14}
\end{equation}
and require the continuity condition. We then get

\begin{equation}
\overline{\psi }[i\gamma ^{\mu }\partial _{\mu }-m)\psi ]+[i\partial _{\mu }%
\overline{\psi }\gamma ^{\mu }+m]\psi =0.  \eqnum{15}
\end{equation}
Since $\overline{\psi }$ and $\psi $ are linearly independent, the arguement
in each of the square brackets must vanish separately, and we get back the
Dirac equation and its adjoint equation with $m$ being the rest mass of the
particle. That is, complex space-time can reproduce the Dirac equation.

\section{Classification of fermion elementary particles}

If no interaction exists between $P_{i}$ and $P_{j}$ in equation (5), the
Lagrangian of the system in GU is a sum of the $L_{j},$ $j=1,\cdots 4,$
where $L_{j}$ is the Lagrangian of $P_{j}$. It is effectively a 4-body
problem, and the total state function $\Psi $ is a product function $\Pi
\Psi _{j}$ , where $\Psi _{j}$ is a free particle Dirac solution for $P_{j}$
with a spin part and an $x_{\mu }$-part. The $x_{\mu }$-part of $\Psi _{j}$
must satisfy equation (7). The solution of equation (8) represents the $%
x_{\mu }$-parts of $\Psi _{2}\Psi _{3}$ . Since $P_{2}$ is the physical
space-time, $\Psi _{2}$ must have good momentum, so there is only one
momentum solution for $\Psi _{2}$, while $\Psi _{3}$ can take two momentum
solutions like those in equation (8). By exchanging the $x_{4}^{I}$ in $%
P_{2} $ with $x_{4}^{R}$ in $P_{3}$, we get $P_{1}$ and $P_{4}$. Therefore,
the pair $P_{1}P_{4}$ is just the same system as the pair $P_{2}P_{3}$ with
the exchange of \ $x_{4}^{I}$ and $x_{4}^{R}$ and the solution is obtained
by the same way. Thus there are $2\otimes 2$ momentum states in $\Psi $ . On
the other hand, there are 4 spin functions in $\Psi $. Since the spin in the
physical space-time is an eigen state in the rest frame and cannot mix with
other internal states. Therefore, the total internal states for a Dirac
particle can be obtained as $2\otimes 2\otimes 2\otimes 2\otimes 2=2\otimes
(1\oplus 3)\otimes (1\oplus 3)$. That is, fermions are grouped into
doublets, then all the doublets are divided into color singlets (leptons)
and color triplets (quarks), then they are further divided into generation
singlets and generation triplets. There are two possibilities:

(A) generation singlet $(\nu _{\tau }^{\ast }\tau ^{\ast })$ N.F. $(t^{\ast
}b^{\ast })$ N.F.

\ \ \ \ \ generation triplet $\;\left[ 
\begin{array}{l}
(\nu _{\tau }\tau ) \\ 
(\nu _{\mu }\mu ) \\ 
(\nu _{e}e)
\end{array}
\begin{array}{l}
(tb) \\ 
(cs) \\ 
(ud)
\end{array}
\right] $

(B) generation triplet $\left[ 
\begin{array}{l}
(\nu _{\tau }^{\ast }\tau ^{\ast })\text{ N.F.} \\ 
(\nu _{\tau }\tau ) \\ 
(\nu _{\mu }\mu )
\end{array}
\begin{array}{l}
(t^{\ast }b^{\ast })\text{ N.F} \\ 
(tb) \\ 
(cs)
\end{array}
\right] $

\ \ \ \ \ generation singlet $(\nu _{e}e)(ud).$

\bigskip

N. F.=not found yet. Take the fermion isospin $\overline{T}$ to be 1/2 and $%
\overline{T}_{3}$ as 1/2 for the left-hand side of the doublets and -1/2 for
the other. Define the fermion hypercharge as SG/N, where N is the
multiplicity of the color multiplet (1 for leptons and 3 for quarks), and SG
is -1 for leptons and 1 for quarks (opposite sign for antiparticles). Then
the electric charge of the particle is given as $Q=\overline{T}_{3}+%
\overline{Y}/2$.

It may be worthwhile to mention that when the momentum of the particle is
zero, states in GU are degenerate. However, since $\pm k/|\pm k|=\pm 1=%
%TCIMACRO{\underset{k\rightarrow 0}{\lim }}%
%BeginExpansion
\mathrel{\mathop{\lim }\limits_{k\rightarrow 0}}%
%EndExpansion
(\pm k/|k|),$ so the degeneracy is removed.

\section{The non-observation of free quarks and a possible new free quark}

In the GU of equation (5), the real space $x_{i}$ is the overlap of $P_{1}$
and $P_{2}$, (Note that the momentum $p_{i}$ in $P_{2}$ is $\partial
L_{2}/\partial \left( \frac{dx_{i}}{dt}\right) $, while that in $P_{1}$ is $%
\partial L_{1}/\partial \left( \frac{dx_{i}}{d\overline{t}}\right) $which is
unphysical). The states in $P_{2}$ do not mix with other states. This can be
regarded as saying that, in the real space $x_{i}$ , there is a physical
particle (from $P_{2}$ ), and a non-physical particle (from $P_{1}$ ), which
are not identical particles. The imaginary space $\overline{x}_{i}$ is the
overlap of \ $P_{3}$ and $P_{4}$. There are two identical particles which
are both non-physical and each has two momentum states and two spin states,
and can couple to singlets and triplets. If the principle of
anti-symmetrization for fermions can be generalized to be valid in the GU,
then in the above list of (A), the leptons of color singlets and generation
triplets are anti-symmetric states and can exist in free particle states,
while the quarks of color triplets and generation triplets are symmetric
states and cannot exist in free particle states (as expected by
experiments). However, if there is no other condition such as ``electric
charge has to be quantized in units of electron charge e''. then there is a
possibility that the free quarks of color triplets and generation singlet in
(A) can exist in nature. On the contrary, free leptons with a color singlet
and a generation singlet can not be observed. Therefore, if the
anti-symmetrization principle works in the greater universe, the above
possibility (B) should be ruled out due to the existence of free electrons.

The author wishes to thank Professors B. Rosenstein and E. Yen for valuable
discussions, and J. Nester for reading the manuscript.

\section{References}

\begin{enumerate}
\item  H.A. Bethe, and R.W. Jackiw, Intermediate quantum mechanics, third
ed. 1986, Addison-Wesley Publishing Co. New York.

\item  P.A.M. Dirac, The principles of quantum mechanics, fourth ed. 1958
Oxford Univ. Press, London.

\item  L. Lin, On the posibility of a complex 4-dimensional space-time
manifold, Gen. Phys./9804016 (1998).
\end{enumerate}

\bigskip

\bigskip 

\end{document}